\documentclass[12pt]{article}
\usepackage[pdftex]{color}
\usepackage{amsmath,amsthm,amssymb}
\usepackage{wasysym}
\usepackage{graphicx}
\usepackage{color}%
\usepackage[onehalfspacing]{setspace}
\usepackage{lipsum}
\usepackage[lmargin=2.5cm,rmargin=2.5cm,tmargin=3cm,bmargin=3cm]{geometry}
\usepackage{multirow}
\usepackage{hyperref}
\hypersetup{hidelinks}
\usepackage{cite}
\usepackage{cleveref}
\usepackage{accents}
\usepackage[ruled,vlined,linesnumbered]{algorithm2e}
\usepackage[numbers,sort&compress]{natbib}
\usepackage{scalerel}
\usepackage{stackengine}
\usepackage[ruled,vlined,linesnumbered]{algorithm2e}
\stackMath
\definecolor{color}{rgb}{0.8500, 0.3250, 0.0980} 

\newcommand\reallywidefrown[1]{%
\stackon[0pt]{#1}{%
\stretchto{%
  \scaleto{%
    \scalerel*[\widthof{#1}]{\mkern-1.5mu\frown\mkern-2mu}%
    {\rule[-\textheight/2]{1ex}{\textheight}}%
  }{\textheight}%
}{0.8ex}}%
}
\usepackage{nomencl}
\makenomenclature
\usepackage{etoolbox}
\renewcommand\nomgroup[1]{%
  \item[\bfseries
  \ifstrequal{#1}{A}{Variables}{%
  \ifstrequal{#1}{B}{Superscripts}{%
  \ifstrequal{#1}{C}{Subscripts}{%
  \ifstrequal{#1}{D}{Abbreviations}{
  \ifstrequal{#1}{E}{Special functions}{
  \ifstrequal{#1}{F}{Operators}{}}}}}}%
]}
\newcommand{\RE}{\mathrm{Re}}

\begin{document}
\begin{center}
{\rm \bf \Large{Electrical contact with dielectric breakdown of interfacial gap}}
\end{center}

\begin{center}
{\bf Yang Xu}$^{\text{ab}}$\footnote{Corresponding author: yang.xu@hfut.edu.cn}, {\bf Yue Wu}$^{\text{c}}$, {\bf Robert L. Jackson}$^{\text{d}}$\\
\end{center}
\begin{flushleft}
{
$^{\text{a}}$School of Mechanical Engineering, Hefei University of Technology, Hefei, 230009, China \\
$^{\text{b}}$Anhui Province Key Laboratory of Digital Design and Manufacturing, Hefei, 230009, China \\
$^{\text{c}}$Shanghai Advanced Research Institute, Chinese Academy of Sciences, Shanghai 201210, China \\
$^{\text{d}}$Department of Mechanical Engineering, Auburn University, AL, 36849, USA 
}
\end{flushleft}

\begin{abstract}
Electrical contact is fundamental to almost every aspect of modern industry, including the fast-growing electric vehicle industry. In metallic contacts in atmospheric conditions, most of the electrical current passes via the micro-junctions formed between two electrodes. The classic electrical contact theory predicts an infinite current density at the circular contact periphery. In the present work, we explore the influence of the dielectric breakdown of air outside the contact area on the electrical contact interface. Incorporating the discharging boundary condition governed by the modified Paschen law, we develop the numerical model as well as two sets of closed-form solutions for low applied voltage cases where two electrodes are in solid-solid contact and complete separation, respectively. For Hertzian contact, the present work theoretically proves that the ignorance of discharge can lead to a singular current density at the contact periphery and an overestimation of the electrical contact resistance. The current density monotonically increases along the radial direction to a finite value at the contact area periphery, followed by a monotonic drop within the discharge zone. The present study serves as a foundation for the modeling of discharging rough surface electrical contact and sheds light on the machine element surface damages caused by the electrical discharge machining.  
\end{abstract}


\newpage

\nomenclature[A]{$h$}{Surface height of the undeformed parabolic indenter}
\nomenclature[A]{$r$}{Radial coordinate}
\nomenclature[A]{$R$}{Radius of curvature at the apex of the parabolic indenter}
\nomenclature[A]{$E$}{Young's modulus}
\nomenclature[A]{$\nu$}{Poisson's ratio}
\nomenclature[A]{$F$}{Compressive normal force}
\nomenclature[A]{$a$}{Radius of contact area}
\nomenclature[A]{$\delta$}{Indentation, $\delta > 0$: contact phase; $\delta \leq 0$: separation phase}
\nomenclature[A]{$E^*$}{Plane strain modulus $1/E^* = (1 - \nu_1^2)/E_1 + (1 - \nu_2^2)/E_2$, where $E_i$ and $\nu_i$, $i = 1, 2$, are Young's modulus and Poisson's ratio of two contacting bodies}
\nomenclature[A]{$g$}{Interfacial gap between the deformed surface and the rigid flat}
\nomenclature[A]{$\rho_1$}{Resistivity of the parabolic surface}
\nomenclature[A]{$\rho_2$}{Resistivity of the rigid flat}
\nomenclature[A]{$V_1$}{Potential applied on the parabolic surface far from the contact}
\nomenclature[A]{$V_2$}{Potential applied on the rigid flat far from the contact}
\nomenclature[A]{$\Delta V$}{Potential drop across the interface far from the contact}
\nomenclature[A]{$z$}{Coordinate on the $z$-axis}
\nomenclature[A]{$J_r^+$}{Current density of the parabolic indenter in $r$ direction}
\nomenclature[A]{$J_z^+$}{Current density of the parabolic indenter in $z$ direction}
\nomenclature[A]{$J_r^-$}{Current density of the rigid flat in $r$ direction}
\nomenclature[A]{$J_z^-$}{Current density of the rigid flat in $z$ direction}
\nomenclature[A]{$\vec{r}$}{Unit vector pointing in the $r$ direction}
\nomenclature[A]{$\vec{z}$}{Unit vector pointing in the $z$ direction}
\nomenclature[A]{$V^+$}{Potential of the parabolic indenter}
\nomenclature[A]{$V^-$}{Potential of the rigid flat}
\nomenclature[A]{$V$}{Potential drop across the interface, $V(r) = V^+(r, z=g(r)) - V^-(r,z = 0)$}
\nomenclature[A]{$J$}{Interfacial current density perpendicular to the interface, $J(r) = J_z^+(r, z = g(r)) = J_z^-(r, z = 0)$}
\nomenclature[A]{$V'$}{Potential, $V' = \Delta V - V$}
\nomenclature[A]{$\rho$}{Composite resistivity, $\rho = \rho_1 + \rho_2$}
\nomenclature[A]{$R_{\text{c}}$}{Electrical contact resistance}
\nomenclature[A]{$c$}{Outer radius of discharge zone}
\nomenclature[A]{$r^*$}{Dimensionless radial coordinate, $r^* = r/a$ at contact phase and $r^* = r/c$ at separation phase}
\nomenclature[A]{$g^*$}{Dimensionless interfacial gap, $g^* = g/\delta$}
\nomenclature[A]{$V^*$}{Dimensionless potential drop, $V^* = V/\Delta V$}
\nomenclature[A]{$V_{\text{b}}$}{Breakdown voltage}
\nomenclature[A]{$A$}{Constant in Paschen law under the ambient pressure, $A = 1.13 \times 10^3$ mm$^{-1}$}
\nomenclature[A]{$B$}{Constant in Paschen law under ambient pressure, $B = 2.74 \times 10^4$ V $\cdot$ mm$^{-1}$}
\nomenclature[A]{$\gamma_{\text{se}}$}{Secondary electron emission coefficient}
\nomenclature[A]{$g_0$}{Interfacial gap associated with the minimum breakdown voltage in the Paschen law}
\nomenclature[A]{$g_{\text{c}}$}{Transitional gap between field emission process and Townsend process portions in the modified Paschen law}
\nomenclature[A]{$K$}{Dielectric strength of air at micro-gaps}
\nomenclature[A]{$\Omega_{\text{con}}$}{Conductive region over the interface}
\nomenclature[A]{$\Omega_{\text{ncon}}$}{Non-conductive region over the interface}
\nomenclature[A]{$r_{\max}$}{Maximum radius of the computational domain}
\nomenclature[A]{$n$}{Number of nodes on the computational domain}
\nomenclature[A]{$\widetilde{V}$}{Potential $\widetilde{V}(r) = V_{\text{b}}(g(r)) - V(r)$}
\nomenclature[A]{${\bf r}$}{Column vector of discrete radial coordinates}
\nomenclature[A]{${\bf \widetilde{V}}$}{Column vector of $\widetilde{V}$ defined over ${\bf r}$}
\nomenclature[A]{${\bf J}$}{Column vector of $J$ defined over ${\bf r}$}
\nomenclature[A]{$I_{\text{con}}$}{Nodal index set associated with the conductive region}
\nomenclature[A]{$I_{\text{ncon}}$}{Nodal index set associated with the non-conductive region}
\nomenclature[A]{${\bf K}$}{Influence coefficient matrix}
\nomenclature[A]{$K_{ij}$}{Influence coefficient, see Eq. \eqref{eq:IC_K}}
\nomenclature[A]{$\sqrt{\langle |\nabla h|^2 \rangle }$}{Root mean square slope of the surface topography}
\nomenclature[A]{$\sqrt{\langle |\nabla h|^2 \rangle }$}{Root mean square slope of the surface topography}
\nomenclature[A]{$\delta^*$}{Dimensionless indentation, $\delta^* = \delta K/\Delta V$}
\nomenclature[A]{$J^*$}{Dimensionless current density, $J^* = \displaystyle{J/\left(\frac{4}{\pi \rho} \sqrt{\frac{K \Delta V}{R}}\right)}$}
\nomenclature[A]{$R_{\text{c}}^*$}{Dimensionless electrical contact resistance, $R_{\text{c}}^* = R_{\text{c}}/\displaystyle{\left( \frac{3 \rho}{8} \sqrt{ \frac{K}{R \Delta V}} \right)}$}
\nomenclature[A]{$d$}{Vertical separation between the rigid flat and the mean level of the undeformed rough surface}
\nomenclature[A]{$d^*$}{Dimensionless $d$, $d^* = d K/\Delta V$}
\nomenclature[A]{$h^*$}{Dimensionless $h$, $h^* = h K/\Delta V$}
\nomenclature[A]{$N$}{Number of asperities over the nominal contact area}
\nomenclature[A]{$\Phi$}{Probability density}
\nomenclature[A]{$\Omega_{\text{c}}$}{Contact region over the interface}
\nomenclature[A]{$\Omega_{\text{nc}}$}{Non-contact region over the interface}
\nomenclature[A]{$\bar{u}_z$}{Surface displacement of a half-space in $z$ direction}
\nomenclature[A]{$\Delta \bar{u}_z$}{Incremental change of the surface displacement of a half-space in $z$ direction in response to $\Delta p$}
\nomenclature[A]{$\Delta p$}{Incremental change of the contact pressure acting on the boundary of a half-space}
\nomenclature[A]{$\Delta \delta$}{Infinitesimally small indentation}
\nomenclature[A]{$K_{\perp}$}{Normal contact stiffness}
\nomenclature[A]{$p$}{Contact pressure}
\nomenclature[D]{LCP }{Linear complementarity problem} 
\nomenclature[D]{ECR }{Electrical contact resistance} 
\nomenclature[D]{EDM }{Electrical discharge machining}
\nomenclature[D]{EIBD}{Electrically-induced bearing damage}
\nomenclature[D]{FFT }{Fast Fourier transform}
\nomenclature[D]{CG  }{Conjugate gradient}
\nomenclature[D]{PIC/MCC}{Particle-in-cell with the Monte Carlo collision}
\nomenclature[E]{$K(m)$}{Complete elliptic integral of the first kind, $K(m) = \int_0^{\pi/2} \left[1 - m^2 \sin^2(\theta)\right]^{-1/2} \text{d}\theta$}
\nomenclature[E]{$E(m)$}{Complete elliptic integral of the second kind, $E(m) = \int_0^{\pi/2} \left[1 - m^2 \sin^2(\theta) \right]^{1/2} \text{d}\theta$}
\nomenclature[F]{$\RE$}{Real part of a complex number}

\printnomenclature

\section{Introduction}
Many applications depend on functioning electrical connects. This includes the growing areas of electric vehicles \cite{hemanth2021hybrid} and alternative energy sources \cite{whittle2013bearing,zuo2022influence}. Specifically, there are many cases such as the battery electrodes/current collector bars contact in lithium-ion batteries \cite{taheri2011investigating}, switches in the electric power industry \cite{holm2013electric}, the lithium metal/solid-state electrolytes contact in all-solid-state lithium metal batteries \cite{zhang2023pressure}, and pantograph-catenary system in high-speed railways \cite{wu2022pantograph}. This includes connectors designed to carry electrical current and also other mechanical components such as gears and bearings carrying the conduction of leakage or unforeseen currents \cite{liu2020review,prashad2002diagnosis}. Predictive models of these components depend on a theoretical framework that is often fundamentally based on single asperity contact resistance models \cite{jackson2023electrical}.

Electrical contact resistance (ECR) is the added resistance that occurs between contacting conductive surfaces in addition to the bulk resistance. In most cases, this refers to the contact of metals. The contact resistance is composed of two main sources: spreading resistance and film resistance. Less conductive oxides and other materials between the surfaces cause film resistance. Since surfaces are practically always rough, the solid contacts are isolated between the peaks or asperities. This results in the electric current (electrons) to contract as they cross between the isolated contacting asperities and then spread again after. The current lines (electron paths) are constricted at the interface so that electrons can only pass through the conductive channel formed by micro-contacts. Therefore, the electrical contact resistance is the outcome of multiple resistors in parallel formed by micro-contacts. The constricted current with high density further causes the local temperature rise (known as the Joule heating), oxidation and melting of the plating layer, energy loss, and even catastrophic fire. 

Holm employed a capacitance calculation method of equipotential surfaces to derive the spreading resistance between two conductive half spaces connected only by a circular conductive region \cite{holm2013electric}. More detailed derivation can be found in Refs. \cite{Llewellyn-Jones57}. Malucci later employed Holm’s method to calculate the spreading resistance for contact between layered or coated surface asperities \cite{malucci2021making}. Nakamura \cite{nakamura1993constriction} also formulated a predictive function for the spreading resistance of square-shaped contact areas. Asperity contact area shapes and sizes can be predicted via contact mechanics theories for single and multiple asperity elastic and elastic-plastic solid contact models \cite{ghaednia2016comprehensive}. For instance, Shah et al. \cite{shah2011electro} showed that a sinusoidal-shaped asperity contact would begin circular in shape but would morph into a square shape. When the asperities are elongated, these contact areas would also be oval or rectangular in shape. In reality, however, the asperities are likely to be much more complicated in shape, as suggested by measurements \cite{xu2018new,xu2020comparison}. Bowden and Williamson \cite{bowden1958electrical} conducted early experiments on electrical contacts and found that the Joule heating in surfaces had a great influence on their softening or failure. Essentially, they found a healing effect that, in many cases, an electrical contact may soften due to Joule heating and increase the contact area in order to carry the load effectively. However, if too much electrical current is applied, the contact area cannot compensate enough, and a thermal run-away occurs, resulting in the contact failing. Electrical load can also accelerate the wear of a contact and effectively increase the roughness of the worn surface \cite{zuo2022influence}. It is also known that contact resistance can vary with the scale or size of the contact areas \cite{jackson2015rough,malucci2005multi}. Once the contact areas approach the micrometer scale or smaller (i.e., the electron mean free path length \cite{wexler1966size}), the nature of the conduction of electrons can change from a continuum diffusive behavior to a quantum ballistic behavior \cite{sharvin1965possible}. However, in the current work, the focus will be on contacts above this length scale, although the gaps between the surfaces at the edge of contact are thin and may allow for significant electron tunneling. 

The rough surface electrical contact problem has been solved either deterministically using numerical models \cite{leidner2010simulation, li2024efficient,wang2024efficient,sui2023modeling,he2024numerical} or analytically based on the electrical contact model with a circular contact area \cite{greenwood1966constriction,greenwood1966contact,kogut2004electrical,jackson2009simplified, yastrebov2015three, ta2021volumetric}. Barber \cite{barber2003bounds} found an analogy between the incremental elastic contact problem and the electrical contact problem, which deduces a linear relation between the interfacial contact stiffness and the contact conductance (reciprocal of ECR). This linear relation is validated by the numerical simulation \cite{li2024efficient}, but it may not be strictly held in cases of elastoplastic contacts \cite{zhai2016interfacial}, especially under high loads. By employing this analogy, the majority of elastic rough surface contact models can be directly utilized to estimate ECR, provided that the interfacial contact stiffness can be determined \cite{ciavarella2004electrical,paggi2011contact,persson2022electric}. In addition, when the isolated asperity contacts grow due to higher pressures and come closer, they can influence each other through the electrical field \cite{malucci2009impact,greenwood1966constriction} and by diverting the current flow from one contact area to another. In other words, each asperity contact will carry part of the electrical current and can be considered as its own constriction from a finite size to the contact area, as modeled by Rosenfeld and Timsit as a cylindrical conductor with a single constricted contact \cite{rosenfeld1981potential}. This effectively lowers the spreading resistance. Following this, Malucci and Ruffino \cite{malucci2008method} also deduced that current density was a critical way to determine the probability of a single asperity contact degrading or failing. Therefore, an electrical contact with many asperity contacts is more resilient and stable. 

The classic electrical contact theory predicts an inverted bell-shaped current density distribution within the contact area with an infinite value at the contact periphery. Like the stress concentration in the flat-end punch contact problem can lead to yielding at the contact periphery, this current concentration suggests that the electrode material may melt at the same spot irrespective of the electrical loads. Malucci \cite{malucci2016single} highlighted that the presence of this non-physical current concentration arises from the assumption that the contact and non-contact areas are not situated on the same plane. According to Malucci \cite{malucci2016single}, the bridge structure (i.e., the three-dimensional geometries of the interfacial gap) adjacent to the contact area can impact how current is distributed at the interface. Malucci \cite{malucci2016single,malucci2017impact,malucci2018effects} found the analytical solutions for current density distribution associated with three distinct interfacial gap geometries, all of which do not have infinite current density at the contact periphery.

In addition to the co-planar boundary conditions adopted in the classic electrical contact theory, the infinite current density at the contact edge may result from neglecting discharge just outside the contact area. The current concentration near the contact edge can result in an intensified electric field nearby, potentially leading to the breakdown of the insulating medium between two electrodes. Therefore, electrons can move through the adjacent conductive gap, resulting in a possibly continuous and finite current density on both sides of the contact edge. As mentioned earlier, contacts can degrade due to the Joule heating from the current flow. However, as will be explored in this work, there may be other modes of degradation, such as the electrical discharge machining (EDM). The phenomenon of discharge is prevalent in the rapidly expanding electric vehicle industry, where powertrains are predominantly electrified by inverters utilizing low-amplitude, high-frequency alternating current \cite{he2020electrical}. For instance, electrically-induced bearing damage (EIBD) dominantly occurs at the rolling element-inner/outer race interface in the electric motor bearing, which is caused by EDM if there is a lack of electrical insulation \cite{janik2024exploring, bond2024influence, chen2020performance}. Jackson et al. \cite{jackson2024statistical} recently developed an electrified mixed lubrication model where the probability of a discharge zone for varying operating conditions and properties can be estimated based on the assumption that the lubricant has a constant dielectric strength.

In order to address the initial concern of whether discharge can cancel the current concentration at the contact edge despite the contact and non-contact regions being co-planar, we will revisit the classic electrical contact theory (without discharging) in Section \ref{sec:classic_theory}. In Section \ref{sec:ECD_num}, the numerical model for discharging electrical contact is developed, along with two analytical solutions for low applied voltage cases. Section \ref{sec:RandD} thoroughly investigates and discusses the impact of discharge on electrical properties at the interface.

\section{Non-discharging electrical contact}\label{sec:classic_theory}
\begin{figure}[h!]
  \centering
  \includegraphics[width=16cm]{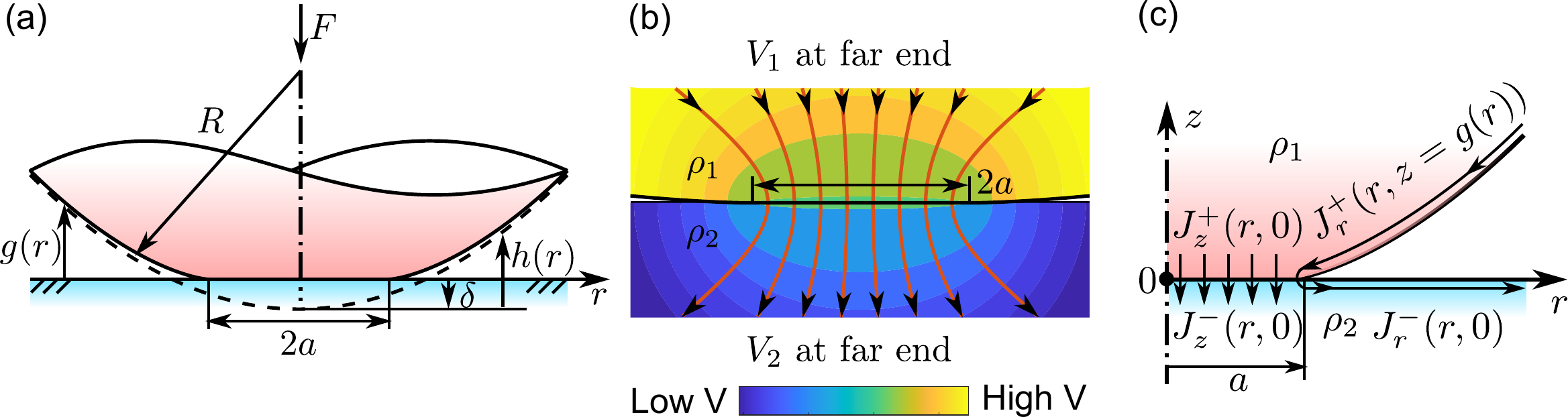}
  \caption{(a) Cross-sectional view of a Hertzian contact. The solid and dashed lines correspond respectively to the deformed and undeformed interface; (b) Cross-sectional view of the potential distribution of an electrical contact at the closed vicinity of the contact area with current lines highlighted; (c) Schematic of current flow at the electrical contact interface (only the right half of the cross-sectional view in (a) is shown).}\label{fig:Fig_1}
\end{figure}

Consider an elastic, axisymmetric, conductive surface in purely normal contact with a rigid, conductive flat (see Fig. \ref{fig:Fig_1}(a)). The undeformed elastic surface has a parabolic shape of $h(r) = r^2/2 R$ where $R$ is the radius of curvature at the apex of the parabolic surface. The substrates beneath the two surfaces are approximately half-spaces. Young's modulus and Poisson's ratio of the elastic surface are $E$ and $\nu$, respectively. The rigid fat is fixed in space, and the parabolic surface is subjected to a compressive normal force $F$. According to Hertzian theory \cite{johnson1987contact}, the purely normal contact between two surfaces results in a circular contact area of radius $a$, 
\begin{equation}
F = \frac{4}{3} E^* R^{-1} a^3,
\end{equation}
and the indentation is
\begin{equation}
\delta = a^2/R, 
\end{equation}
where $E^* = E/(1 - \nu^2)$ is the plane strain modulus. The interfacial gap outside the contact area is 
\begin{equation}\label{eq:gap}
g(r > a) = \frac{a}{\pi R} \sqrt{r^2 - a^2} + \frac{2 a^2 - r^2}{R} \left[ \frac{1}{\pi} \arcsin \left( \frac{a}{r} \right) - \frac{1}{2} \right], 
\end{equation}
and is zero elsewhere. We assume $a \ll R$ so that the surface gradient ($\mathrm{d} g/\mathrm{d} r$) at the vicinity of the contact area is nearly zero (i.e., the small slope assumption). 

Let the potential boundaries of the two contacting bodies far from the contact be $V_1$ and $V_2 (< V_1)$. The resistivities of the two half-spaces are $\rho_1$ and $\rho_2$, respectively. The current is driven by this potential difference ($\Delta V = V_1 - V_2$), and all current lines are constricted over the circular contact area if dielectric breakdown does not occur at the interface. A graphical illustration of this electrical contact problem can be found in Fig. \ref{fig:Fig_1}(b). The current density vectors ($J_r^+(r, z \geq g(r)) \vec{r} + J_z^+(r, z \geq g(r)) \vec{z}$ and $J_r^-(r, z \leq 0) \vec{r} + J_z^-(r, z \leq 0) \vec{z}$) and potential ($V^+(r, z \geq g(r))$ and $V^-(r, z \leq 0)$) of both contacting bodies can be solved either analytically using the Hankel transform \cite{Llewellyn-Jones57} or numerically using the conjugate gradient (CG) method and the fast Fourier transform (FFT) \cite{wang2024efficient}. 

Let the potential drop across the interface be $V(r) = V^+(r, z = g(r)) - V^-(r, z = 0)$. Since current flows across the contact area perpendicular to the interface (see Fig. \ref{fig:Fig_1}(c)), according to the definition of potential, we have $V(r \leq a) = \rho_1 \int_{0^+}^{0} J_z^+(r, z')\text{d}z' + \rho_2 \int_{0}^{0^-}J_z^-(r, z') \text{d}z'$ = 0. On the contrary, current cannot pass through the gap outside the contact area if the gas medium is a perfect dielectric with an infinite breakdown voltage. The current outside the contact area first flows toward the contact edge ($r = a$) parallel to the indenter's surface and then flows back in the same fashion on the rigid, flat surface toward the outer radius (see Fig. \ref{fig:Fig_1}(c)). Based on the small slope assumption, we can formulate the potential drop approximately as $V(r) = \rho_1 \int_a^r J_r^+(r', z = g(r')) \text{d}r' + \rho_2 \int_a^r J_r^-(r', z = 0) \text{d} r' > 0$. The potential drop $V(r)$ gradually increases with $r$ and eventually converges to $\Delta V$. Let the interfacial current density flowing perpendicular to the interface be $J(r) = J_z^+(r, z = g(r)) = J_z^-(r, z = 0)$, where $J(r \leq a) > 0$ and $J(r > a) = 0$. We can analytically solve $J(r)$ and $V(r)$ based on Barber's elastic-electrical analogy \cite{barber2003bounds}; the detailed derivation can be found in Appendix A. The closed-form solutions of non-zero $J(r)$ and $V(r)$ are

\begin{align}
J(r \leq a) &= \frac{2 \Delta V}{\pi \rho} (a^2 - r^2)^{-1/2},  \label{eq:J_closeform} \\
V(r > a) &= \Delta V \left[ 1 - \frac{2}{\pi} \arcsin \left(\frac{a}{r}\right) \right], \label{eq:V_closeform}
\end{align}
where $\rho = \rho_1 + \rho_2$. As far as authors know, this is the first time that the closed form of $V(r > a)$ is explicitly given. It is surprising to see that the voltage drop, $V(r)$, given in Eq. \eqref{eq:V_closeform} is independent of the resistivities of the two electrodes and can be entirely determined based on the geometric properties of the electrical contact. Similarly, the expression of the interfacial gap, $g(r > a)$, in Eq. \eqref{eq:gap} does not explicitly rely on the mechanical properties of the two electrodes, whose effect on $g(r)$, however, is indirectly introduced through $a$. The electrical contact resistance is \cite{holm2013electric,greenwood1966contact,Llewellyn-Jones57}
\begin{equation}\label{eq:Rc_Holm}
R_{\text{c}} = \Delta V/\int_0^a J(r) 2 \pi r \text{d}r = \frac{\rho}{4a}. 
\end{equation}

For an axisymmetric electrical contact, $V(r)$ and $J(r)$ can also be solved numerically from the Karush-Kuhn-Tucker condition: 

\begin{alignat}{3}
&V(r) = 0, ~~~ && J(r) > 0, ~~~ && r \leq a, \label{eq:VJ_contact} \\
&V(r) > 0, ~~~ && J(r) = 0, ~~~ && r > a. \label{eq:VJ_noncontact}  
\end{alignat}
The radial distribution of the potential drop, $V(r)$, can be deduced from Eq. \eqref{eq:Vp},
\begin{equation}\label{eq:V_1D}
  V(r) = \Delta V - \frac{2 \rho}{\pi} \int_0^a J(s) K\left(m = \frac{2 \sqrt{r s}}{r + s} \right) \frac{s}{r + s} \text{d}s,
\end{equation}
where $K(m) = \int_0^{\pi/2} \left[ 1 - m^2 \sin^2(\theta)\right]^{-1/2} \text{d} \theta$ is the complete elliptic integral of the first kind. The discretized forms of Eqs. (\ref{eq:VJ_contact}-\ref{eq:V_1D}) result in a linear complementarity problem (LCP), which can be iteratively solved by the CG method \cite{sui2023modeling, li2024efficient, wang2024efficient}. The details of the numerical model can be found in Section \ref{subsec:numerical_model}. 

\section{Discharging electrical contact}\label{sec:ECD_num}

Let us define three dimensionless variables, $r^* = r/a$, $g^* = g/\delta$ and $V^* = V/\Delta V$. The dimensionless forms of Eqs. \eqref{eq:gap} and \eqref{eq:V_closeform} are written as
\begin{align}
g^*(r^* > 1) &= \frac{1}{\pi} \sqrt{r^{*2} - 1} + (2 - r^{*2}) \left[ \frac{1}{\pi} \arcsin(1/r^*) - \frac{1}{2} \right], \label{eq:gap_dimensionless} \\
1/r^* &= \sin\left(\frac{\pi}{2}(1 - V^*)\right). \label{eq:V_dimensionless}
\end{align} 
Substituting Eq. \eqref{eq:V_dimensionless} into Eq. \eqref{eq:gap_dimensionless}, we can obtain the dimensionless relation $g^*(V^*)$ in a closed form:
\begin{equation}\label{eq:Vg_nodischarge}
g^*(V^*) = \frac{1}{2} V^* \cot^2\left( \frac{\pi}{2} (1 - V^*) \right) + \frac{1}{\pi} \cot \left( \frac{\pi}{2} (1 - V^*) \right) - \frac{1}{2} V^*,
\end{equation}
where $V^* \in [0, 1]$. As $g^* \to 0^+$, $\text{d}V^*/\text{d} g^* \to \infty$. This singular behavior implies that the electric field at the opening edge adjacent to the contact area is infinite. Therefore, the gas medium with a finite electrical strength inevitably breaks down. The conductive channel shall initiate either at the contact edge when two surfaces are in solid-solid contact or at the minimum gap location ($r=0$) when two surfaces are completely separated. As suggested by one of the reviewers, when two electrodes are in adhesive contact in the JKR limit \cite{johnson1971surface}, $\text{d}V^*/\text{d} g^*$ becomes finite at the contact edge and decreases with the increasing $r^*$. In that case, the dielectric breakdown of air may not occur over the interface.

\subsection{Air breakdown}

\begin{figure}[h!]
  \centering
  \includegraphics[width=8cm]{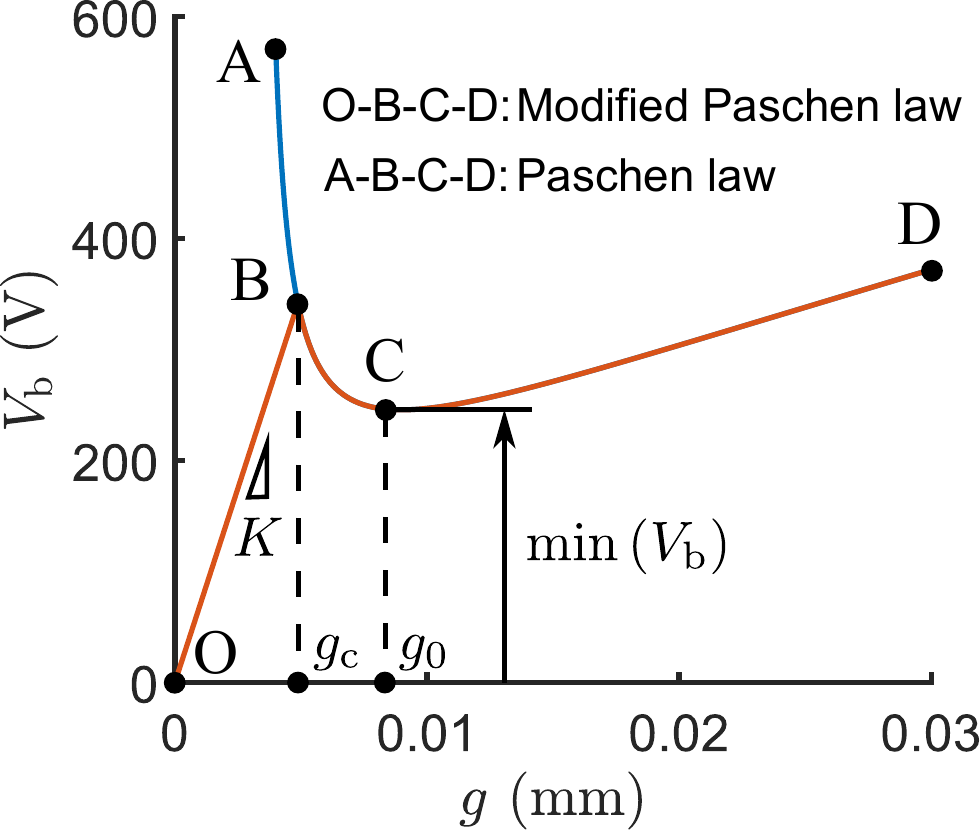}
  \caption{Variations of $V_{\text{b}}$ with $g$ governed by the Paschen law (Eq. \eqref{eq:Paschen_curve}) and the modified Paschen law (Eq. \eqref{eq:modified_Paschen}) under the ambient pressure of $1 \times 10^5$ Pa.}\label{fig:Fig_2}
\end{figure}

Let us assume that the gas medium between two electrodes is air with no water molecules, dust, or other contaminants. For higher separations ($g > 20$ $\mu$m), air breakdown between electrodes follows the Townsend process, which is composed of electron impact ionization and secondary electron emission \cite{go2014microscale}. Let the breakdown voltage be $V_{\text{b}}$, beyond which air breakdown occurs. Under the ambient pressure of $1 \times 10^5$ Pa, the $V_{\text{b}}(g)$ relation can be characterized by the Paschen's law \cite{husain1982analysis}, 
\begin{equation}\label{eq:Paschen_curve}
V_{\text{b}} = \frac{B \cdot g}{\ln(A \cdot g) - \ln(\ln(1 + 1/\gamma_{\text{se}}))},
\end{equation}
where $A = 1.13 \times 10^3$ mm$^{-1}$, $B = 2.74 \times 10^{4}$ V $\cdot$ mm$^{-1}$, and $\gamma_{\text{se}}$ is the secondary electron emission coefficient, which is closely related to the electrode material. Eq. \eqref{eq:Paschen_curve} has a minimum voltage at $g_0 = e \ln(1 + 1/\gamma_{\text{se}})/A$. Assuming two electrodes are copper ($\gamma_{\text{se}} = 0.025$), then $\min(V_{\text{b}}) \approx 244.8$ V at $g_0 \approx 9.0$ $\mu$m. The corresponding Paschen law is shown in Fig. \ref{fig:Fig_2}, which consists of two branches on both sides of the minimum voltage point (C in Fig. \ref{fig:Fig_2}). In the right branch ($\reallywidefrown{CD}$), $V_{\text{b}}$ increases monotonically with $g$. The left branch ($\reallywidefrown{ABC}$) fails to predict the breakdown voltage at lower gap ranges because field emission from the cathode electrode induced by the high electric field in the micro-/nano-scale gaps is not included in the Paschen law. Many micro-/nano-scale air gap breakdown tests \cite{torres1999electric,lee2001arc,slade2002electrical} conducted in the ambient environment have shown a linear relation of $V_{\text{b}}(g)$ of which the slope (dielectric strength) $K \in \left[ 6.5 \times 10^{4}, 1.1 \times 10^{5} \right]$ V/mm. In the present work, we propose a modified Paschen law
\begin{equation}\label{eq:modified_Paschen}
V_{\text{b}} = 
\begin{cases}
K \cdot g ~~~ g \leq g_{\text{c}}, \\
\displaystyle{\frac{B \cdot g}{\ln(A \cdot g) - \ln\left( \ln \left(1 + 1/\gamma_{\text{se}}\right) \right)}} ~~~ g > g_{\text{c}},
\end{cases}
\end{equation}
where $K = 7 \times 10^4$ V/mm in the present work and transitional gap $g_{\text{c}} = \ln(1 + 1/\gamma_{\text{se}}) e^{B/K}/A \approx 4.9$ $\mu$m.

In the discharge zone at the interface, the $V(g)$ relation should strictly follow the Paschen law, i.e., $V(g) = V_{\rm b}(g)$. With such a complex boundary condition, it is nearly impossible to completely solve the discharging electrical contact problem analytically. Thus, we formulate the numerical model first in Section \ref{subsec:numerical_model}. Then, two solutions for low voltage applications are respectively derived in closed forms in Section \ref{subsec:analalytical_solu} when two electrodes are in solid-solid contact and complete separation. 

\subsection{Numerical model}\label{subsec:numerical_model}
Let us consider the electrical contact problem shown in Fig. \ref{fig:Fig_1}, where discharge events between two electrodes are governed by the modified Paschen law given in Eq. \eqref{eq:modified_Paschen}. First, we define $\Omega_{\text{con}}$ and $\Omega_{\text{ncon}}$ as the conductive and non-conductive regions, respectively. The conductive region is composed of the contact area and part of the non-contact area where discharge occurs. Then, we define a new potential variable $\widetilde{V}(r) = V_{\text{b}}(g(r)) - V(r)$. Finally, the discharging electrical contact problem can be formulated as follows:

\begin{alignat}{3}
&\widetilde{V}(r) = 0, ~~~ && J(r) > 0, ~~~ && r \in \Omega_{\text{con}}, \label{eq:VJ_conductive} \\
&\widetilde{V}(r) > 0, ~~~ && J(r) = 0, ~~~ && r \in \Omega_{\text{ncon}}, \label{eq:VJ_nonconductive}  
\end{alignat}
where the radial distribution $\widetilde{V}(r)$ is deduced from Eq. \eqref{eq:V_1D}, i.e., 
\begin{equation}\label{eq:V_tilde_1D}
\widetilde{V}(r) = V_{\text{b}}(g(r)) - \Delta V + \frac{2 \rho}{\pi} \int_0^{\infty} J(s) K\left(m = \frac{2 \sqrt{r s}}{r + s} \right) \frac{s}{r + s} \text{d}s.
\end{equation}
Let us uniformly discretize a finite domain ($r \in [0, r_{\max}]$) into a column vector ${\bf r} = [r_1, r_2, \cdots, r_n]^T$, where $n$ is the number of nodes. The radial distribution $g(r)$, $\widetilde{V}(r)$, and $J(r)$ defined over ${\bf r}$ are ${\bf g}$, ${\bf \widetilde{V}}$, and ${\bf J}$, respectively. The discretized form of Eqs. (\ref{eq:VJ_conductive}--\ref{eq:V_tilde_1D}) are 
\begin{alignat}{3}
\widetilde{V}_i &= 0, ~~~ J_i > 0, ~~~ i \in I_{\text{con}}, \label{eq:LCP_1} \\
\widetilde{V}_i &> 0, ~~~ J_i = 0, ~~~ i \in I_{\text{ncon}}, \label{eq:LCP_2} \\
\widetilde{V}_i &= V_{\text{b}}(g_i) - \Delta V + \sum \limits_{j = 1}^{n-1} K_{ij} J_j, ~~~ \forall i, \label{eq:LCP_3} 
\end{alignat}
where $I_{\text{con}} = \{i |J_i > 0, i = 1, \cdots, n \}$, $I_{\text{ncon}}  = \{i |J_i = 0, i = 1, \cdots, n \}$, and $I_{\text{con}} \bigcap I_{\text{ncon}} = \varnothing$, $K_{ij}$ is the coefficient of the influence matrix ${\bf K}$ and can be determined analytically using Riemann-Stieltjes integral as follows \cite{xu2024adhesion}:
\begin{equation}\label{eq:IC_K}
K_{ij} = 
\begin{cases}
\rho (r_{j+1} - r_{j}) ~~~ i = 1, \\
\rho \left[ G(r_{j+1}/r_i) - G(r_{j}/r_i) \right] r_i  ~~~ 2 \leq i \leq n.
\end{cases}
\end{equation}
The function $G(k)$ has the following closed-form expression:
\begin{equation}
G(k) = \frac{1}{\pi} \left[ (1 + k) E\left( \frac{2 \sqrt{k}}{1 + k} \right) - (1 - k) K\left( \frac{2 \sqrt{k}}{1 + k} \right) \right],
\end{equation}
where $E(m) = \int_0^{\pi/2} \left[1 - m^2 \sin^2(\theta) \right]^{1/2} \text{d}\theta$ is the complete elliptic integral of the second kind. Eqs. (\ref{eq:LCP_1}--\ref{eq:LCP_3}) forms the LCP, which can be effectively solved by many classic optimization solvers \cite{polonsky1999numerical,bemporad2015optimization,xi2017linear}. 
In the present work, the CG method is used to iteratively solve the above LCP. The algorithm, inspired by a similar one proposed by Polonsky and Keer \cite{polonsky1999numerical} for the contact mechanics problem, is given below and is implemented in Matlab. Due to the high current density gradient at the vicinity of $r/a = 1$, the corresponding mesh density should be sufficiently high to guarantee the smoothness of $J(r)$. The above LCP can be easily adapted to formulate a general three-dimensional discharging electrical contact problem between a rigid flat and an elastic half-space \cite{wang2024efficient} with an arbitrary three-dimensional shaped boundary $h(x, y)$, where the root mean square slope $\sqrt{\langle |\nabla h|^2 \rangle} \ll 1$. The interfacial gap between the interacting surfaces can be formulated by an alternative LCP and solved numerically by the FFT-accelerated CG method \cite{polonsky1999numerical}.

\begin{algorithm}[h!]\label{alg:CG}
\setstretch{1}
\SetKwInOut{Input}{input}
\SetKwInOut{Output}{output}
\caption{Polonsky \& Keer model (1999)\\
The algorithm for solving Eqs. (\ref{eq:LCP_1}-\ref{eq:LCP_3}) using the CG method developed by Polonsky and Keer \cite{polonsky1999numerical} with minor changes.}
\Input{${\bf K}$, ${\bf g}$, $I_{\text{c}}$, $\Delta V$, $\epsilon_0$.}
\Output{${\bf J}$, ${\bf V}$.}
Initialize $\epsilon = 1$, ${\bf V}_{\text{b}} = V_{\text{b}}({\bf g})$, $I_{\text{con}} = I_{\text{c}}$, $J_{i \in I_{\text{con}}} = \Delta V$, $J_{i \notin I_{\text{con}}} = 0$, ${\bf J}^{\text{old}} = {\bf J}$\;
\While{$\epsilon > \epsilon_0$}{
  $\widetilde{{\bf V}} = {\bf V}_{\text{b}} - \Delta V + {\bf K} \cdot {\bf J}$ \tcp*{Eq. \eqref{eq:LCP_3}}
  ${\bf t} = \text{zeros}(n, 1)$\;
  $t_{i \in I_{\text{con}}} = \widetilde{V}_{i \in I_{\text{con}}}$ \tcp*{Calculate the conjugate direction}
  ${\bf r}' = {\bf K} \cdot {\bf t}$\;
  $\tau = \sum \limits_{i \in I_{\text{con}}} \widetilde{V}_i t_i/\sum \limits_{i \in I_{\text{con}}} r'_i t_i$ \tcp*{Calculate the step size}

  ${\bf J} \leftarrow {\bf J} - \tau {\bf t}$ \tcp*{Correct ${\bf J}$ along the conjugate direction}

  {\bf Set} all negative $J_i$ to zero\;

  {\bf Update} $I_{\text{ol}} = \{i|J_i = 0, \widetilde{V}_i < 0, i = 1, \cdots, n \}$\;

  $J_{i \in I_{\text{ol}}} \leftarrow J_{i \in I_{\text{ol}}} - \tau \widetilde{V}_{i \in I_{\text{ol}}}$\tcp*{Correct negative $\widetilde{V}_i$}

  $\epsilon = \sqrt{\sum \limits_{i = 1}^n (J_i - J_i^{\text {old}})^2}/\sqrt{\sum \limits_{i = 1}^n J_i^2}$ \tcp*{Error}
  ${\bf J}^{\text{old}} = {\bf J}$, $I_{\text{con}} = \{ i |J_i > 0, i = 1, \cdots, n \}$\;
 }
${\bf V} = {\bf V}_{\text{b}} - \widetilde{{\bf V}}$.
\end{algorithm}

When two electrodes are completely separated ($a = 0$ and $\delta = -g(r = 0) \leq 0$), the governing equations of electrodynamics switch from Ampere's law to Gauss' law. In the latter electrostatic analysis, the permittivity of the dielectric media between two electrodes (for example, air and lubricant) greatly influences the local electric field, which is essential to the outburst of discharge events \cite{mahdy1998electrode}. When the interfacial gap is negligibly small compared with the electrode dimension, we can assume that the interfacial gap at the discharge zone is ``artificially" closed. Therefore, Eqs. (\ref{eq:VJ_conductive}--\ref{eq:V_tilde_1D}) remain applicable when $\delta \leq 0$ if air breakdown occurs.

\subsection{Electrical contact with discharging - analytical model}\label{subsec:analalytical_solu}
The accuracy of the numerical model developed in Section \ref{subsec:numerical_model} is highly sensitive to the local mesh density. This mesh-dependent feature complicates its usage unless an adaptive mesh is used so that the element size is adaptively refined at a high current density gradient region. If we restrict the electrical contact problems to those low applied voltage cases, where $\Delta V \ll \min(V_{\text{b}}) \approx 244.8$ V, we can obtain two sets of closed-form solutions, respectively, when two surfaces are in solid-solid contact (contact phase) and complete separation (separation phase).

\subsubsection{Contact phase}
The following derivation is inspired by the double Hertzian solution proposed by Greenwood and Johnson \cite{greenwood1998alternative} for adhesive contact. Consider a Hertzian pressure-like current density distribution 
\begin{equation}
J(r) = J_0 \cdot \RE \sqrt{1 - r^2/a^2},
\end{equation}
where $\RE$ represents the real part of a complex variable. According to the analogy between integral forms of $\bar{u}_z$ (surface normal displacement, Eq. \eqref{eq:uz_bar}) and $V' = \Delta V - V$ (Eq. \eqref{eq:Vp}), the resultant potential distribution can be directly obtained from Hertzian theory \cite{johnson1987contact},

\begin{align}
V'(r \leq a) &= \displaystyle{\frac{\pi \rho J_0}{8 a}} (2 a^2 - r^2), \label{eq:Auxiliary_sol_1}\\
V'(r > a) &= \displaystyle{\frac{\rho J_0}{4 a}} \left[(2 a^2 - r^2) \text{arcsin}(a/r) + a \sqrt{r^2 - a^2} \right]. \label{eq:Auxiliary_sol_2}
\end{align}
Eqs. \eqref{eq:Auxiliary_sol_1} and \eqref{eq:Auxiliary_sol_2} are not the final solutions but rather auxiliary solutions of $V'(r)$ due to a Hertzian pressure-like current density distribution. These auxiliary solutions can serve as building blocks to achieve the final closed-form solutions.

Let us provide an initial guess of $J(r)$ for the present problem as follows:
\begin{equation}\label{eq:Jr_guess}
J(r) = J_1 \cdot \RE \sqrt{1 - r^2/c^2} - J_0 \cdot \RE \sqrt{1 - r^2/a^2},
\end{equation}
where $r \in [a, c]$ and $c$ is the outer radius of the discharge zone. The potential distribution within the contact area is the superposition of two auxiliary solutions,
\begin{equation}\label{eq:V_res1}
V'(r \leq a)= \frac{\pi \rho}{4} (J_1 c - J_0 a) + \frac{\pi \rho}{8} \left( J_0/a - J_1/c \right) r^2. 
\end{equation}
Let $J_1 = J_0 c/a$, then the potential within the contact area deduces to a constant distribution, 
\begin{equation}
V'(r \leq a) = \frac{\pi \rho J_0}{4 a} (c^2 - a^2).
\end{equation}
According to the boundary condition $V'(r \leq a) = \Delta V$, we can establish the relationship between two unknowns ($J_0$ and $c$) as follows:
\begin{equation}\label{eq:V_cont}
J_0 = \frac{4 a \Delta V}{\pi \rho (c^2 - a^2)}. 
\end{equation}
Eq. \eqref{eq:V_cont} guarantees the zero voltage drop condition within the contact area (i.e., $V(r \leq a) = 0$). Substituting Eq. \eqref{eq:V_cont} into Eq. \eqref{eq:Jr_guess}, we can get the final form of $J(r)$ as
\begin{equation}
J(r) = \frac{4 a \Delta V}{\pi \rho (c^2 - a^2)} \left[ \RE \sqrt{c^2/a^2 - r^2/a^2} - \RE \sqrt{1 - r^2/a^2} \right].
\end{equation}
The final form of $V(r) = \Delta V - V'(r)$ is 

\begin{alignat}{3}
V(r \in (a, c]) &= \frac{R \Delta V }{c^2 - a^2} g(r), \label{eq:Vr_inside}\\
V(r > c) &= \Delta V  - \frac{\Delta V }{\pi (c^2 - a^2)} \bigg\{&&\left[(2 c^2 - r^2) \text{arcsin}(c/r) + c \sqrt{r^2 - c^2} \right] - \notag \\
& ~~~~ &&\left[(2 a^2 - r^2) \text{arcsin}(a/r) + a \sqrt{r^2 - a^2} \right] \bigg\},
\end{alignat}
and $V(r \leq a) = 0$. Surprisingly, we find that $V(r \in (a, c])$ is a linear function of $g(r \in (a, c])$ given in Eq. \eqref{eq:gap}. In low voltage applications, discharge is mainly due to the field emission of the cathode so that the boundary condition within $r \in [a, c]$ is $V(r) = V_{\text{b}}(g(r)) = K g(r)$. Therefore, the unknown $c$ can be determined analytically from Eq. \eqref{eq:Vr_inside} as
\begin{equation}\label{eq:c_ana}
c/a = \sqrt{\frac{R \Delta V}{K a^2} + 1}.
\end{equation}
Finally, the contact resistance is 
\begin{equation}\label{eq:Rc_contact}
R_{\rm{c}} = \frac{3 (c^2 - a^2)}{8 \left(c^3 - a^3\right)}\rho.
\end{equation}

\subsubsection{Separation phase}
When two surfaces are completely separated with a positive minimum gap ($\delta < 0$), the interfacial gap between the undeformed parabolic surface and the rigid flat is $g(r) = -\delta + r^2/2R$. The current can still pass through the central gap ($r \leq c$) as long as $\Delta V + K \delta > 0$. The discharge zone satisfies the potential boundary condition: $V(r) = V_\text{b}(g(r)) = K g(r)$. Therefore, we can write the mixed boundary conditions as 

\begin{alignat}{3}
J(r) > 0&, ~~~V'(r)&& = (\Delta V + K             \delta) - K r^2/2 R, ~~~ &&r \leq c, \label{eq:SP_JV_dis}\\
J(r) = 0&, ~~~V'(r)&& > 0, ~~~ &&r > c. \label{eq:SP_JV_nondis}
\end{alignat}
Eqs. \eqref{eq:SP_JV_dis} and \eqref{eq:SP_JV_nondis} are similar to the mixed boundary conditions of Hertzian contact. Thus, $J(r)$ must have the same form as that of Hertzian contact pressure \cite{johnson1987contact}, 
\begin{equation}
J(r) = J_{\max} \cdot \RE \sqrt{1 - r^2/c^2}.
\end{equation}
According to Eq. \eqref{eq:Auxiliary_sol_1}, the corresponding potential distribution inside the discharge zone is
\begin{equation}\label{eq:V_SP_2}
V'(r \leq c) = \frac{\pi}{4} \rho c J_{\max} \left( 1 - r^2/2 c^2 \right).
\end{equation}
By setting the right side of Eq. \eqref{eq:V_SP_2} equal to the right side of $V'(r)$ in Eq. \eqref{eq:SP_JV_dis}, we can solve for two unknowns, 
\begin{equation}
c = \sqrt{R (\Delta V + K \delta)/K}, ~~~ J_{\max} = \frac{4}{\pi \rho} \sqrt{(\Delta V + K \delta) K/R}.
\end{equation}
The final forms of $V(r)$ and $J(r)$ are given below:

\begin{align}
V(r \leq c) &= K g(r), \\
V(r > c) &= \Delta V - \frac{K}{\pi R} \left[ (2 c^2 - r^2) \text{arcsin}(c/r) + c \sqrt{r^2 - c^2} \right], \\
J(r) &= \frac{4}{\pi \rho} \sqrt{(\Delta V + K \delta) K/R} \cdot \RE \sqrt{1 - r^2/c^2}.
\end{align}
The electrical contact resistance is 
\begin{equation}\label{eq:Rc_separation}
R_{\rm{c}} = \frac{3 \Delta V \rho}{8 \left(\Delta V + K \delta \right) c}.
\end{equation}
The continuity of the electrical contact resistance at $\delta = 0$ can be easily checked by letting $a = 0$ and $\delta = 0$ in Eqs. \eqref{eq:Rc_contact} and \eqref{eq:Rc_separation} so that we have
\begin{equation}
R_{\rm{c}}|_{\delta = 0} = \frac{3 \rho}{8 c}.
\end{equation} 

\subsubsection{Dimensionless forms}\label{subsec:dimensionless_form}
In this section, we summarize all analytical solutions in their dimensionless forms. Let us first propose the following dimensionless group: 

\begin{align}
V^* &= V/\Delta V, ~~~ J^* = J/\left(\frac{4}{\pi \rho} \sqrt{\frac{K \Delta V}{R}} \right), ~~~ R_{\text{c}}^* = R_{\text{c}}/ \left( \frac{3 \rho}{8} \sqrt{\frac{K}{R \Delta V}} \right), \notag \\
\delta^* &= \delta K/\Delta V, ~~~ r^* = r/a~(\text{contact~phase})~\text{or}~r^* = r/c~(\text{separation~phase}).  
\end{align}
Then, all non-trivial electrical interfacial solutions in dimensionless forms are tabulated below:
\begin{itemize}
\item {\bf Non-discharging state} ($\delta^* > 0$)
\begin{subequations} 

\begin{align}
V^*(r^* > 1) &= 1 - \frac{2}{\pi} \text{arcsin}(1/r^*), \label{eq:V_Holm_dless}\\
J^*(r^*) &= \frac{1}{2} (\delta^*)^{-1/2} /\text{Re} \sqrt{1 - r^{*2}}, \label{eq:J_Holm_dless} \\
R_c^* &= \frac{2}{3} (\delta^*)^{-1/2}. \label{eq:Rc_Holm_dless}
\end{align}
\end{subequations}
\item {\bf Discharging state at contact phase} ($\delta^* > 0$)
\begin{subequations}

\begin{align}
V^*(r^* \in (1, c^*]) =& \delta^* \left\{ \frac{1}{\pi} \sqrt{r^{*2} - 1} + (2 - r^{*2}) \left[ \frac{1}{\pi} \text{arcsin}(1/r^*) - \frac{1}{2} \right] \right\}, \label{eq:V_dis_cont_dless1}\\
V^*(r^* > c^*) =& 1 - \frac{\delta^* }{\pi} \bigg\{\left[(2c^{*2} - r^{*2})\text{arcsin}(c^*/r^*) + c^* \sqrt{r^{*2} - c^{*2}} \right] - \notag \\
&\left[(2 - r^{*2})\text{arcsin}(1/r^*) + \sqrt{r^{*2} - 1} \right]\bigg\}, \label{eq:V_dis_cont_dless2} \\
J^*(r^*) =& \sqrt{\delta^*} \left[ \text{Re} \sqrt{c^{*2} - r^{*2}} - \text{Re} \sqrt{1 - r^{*2}} \right], \label{eq:J_dis_cont_dless} \\
R_{\rm{c}}^* =& \left[(1 + \delta^*)^{3/2} - (\delta^*)^{3/2}\right]^{-1}, \label{eq:Rc_dis_cont_dless}
\end{align}
\end{subequations}
where $c^* = \sqrt{1 + 1/\delta^*}$. By setting the right side of Eq. \eqref{eq:Rc_Holm_dless} equal to the right side of Eq. \eqref{eq:Rc_dis_cont_dless}, we find no real root for $\delta^*$. Thus, it is easy to prove that discharging $R_{\text{c}}^*$ is always less than non-discharging $R_{\text{c}}^*$, $\forall \delta^* \geq 0$. A graphical illustration of this inequality can be found in Fig. \ref{fig:Fig_R4}.  

\item {\bf Discharging state at separation phase} ($\delta^* \leq 0$)
\begin{subequations}

\begin{align}
V^*(r^* \leq 1) &= -\delta^* + \frac{1}{2} (1 + \delta^*) (r^*)^2, \label{eq:V_dis_sep_dless1} \\
V^*(r^* > 1) &= 1 - \frac{1}{\pi} ( 1 + \delta^* ) \left[ (2 - r^{*2}) \arcsin(1/r^*) + \sqrt{r^{*2} - 1} \right], \label{eq:V_dis_sep_dless2} \\
J^*(r^*) &= \sqrt{1 + \delta^*} \text{Re} \sqrt{1 - r^{*2}}, \label{eq:J_dis_sep_dless} \\
R_{\text{c}}^* &= (1 + \delta^*)^{-3/2}. \label{eq:Rc_dis_sep_dless}
\end{align}
\end{subequations}
\end{itemize}
For a fixed $r^*$, all dimensionless results shown in this section only depend on $\delta^*$. 

\section{Results and discussion}\label{sec:RandD}

\begin{figure}[h!]
  \centering
  \includegraphics[width=16cm]{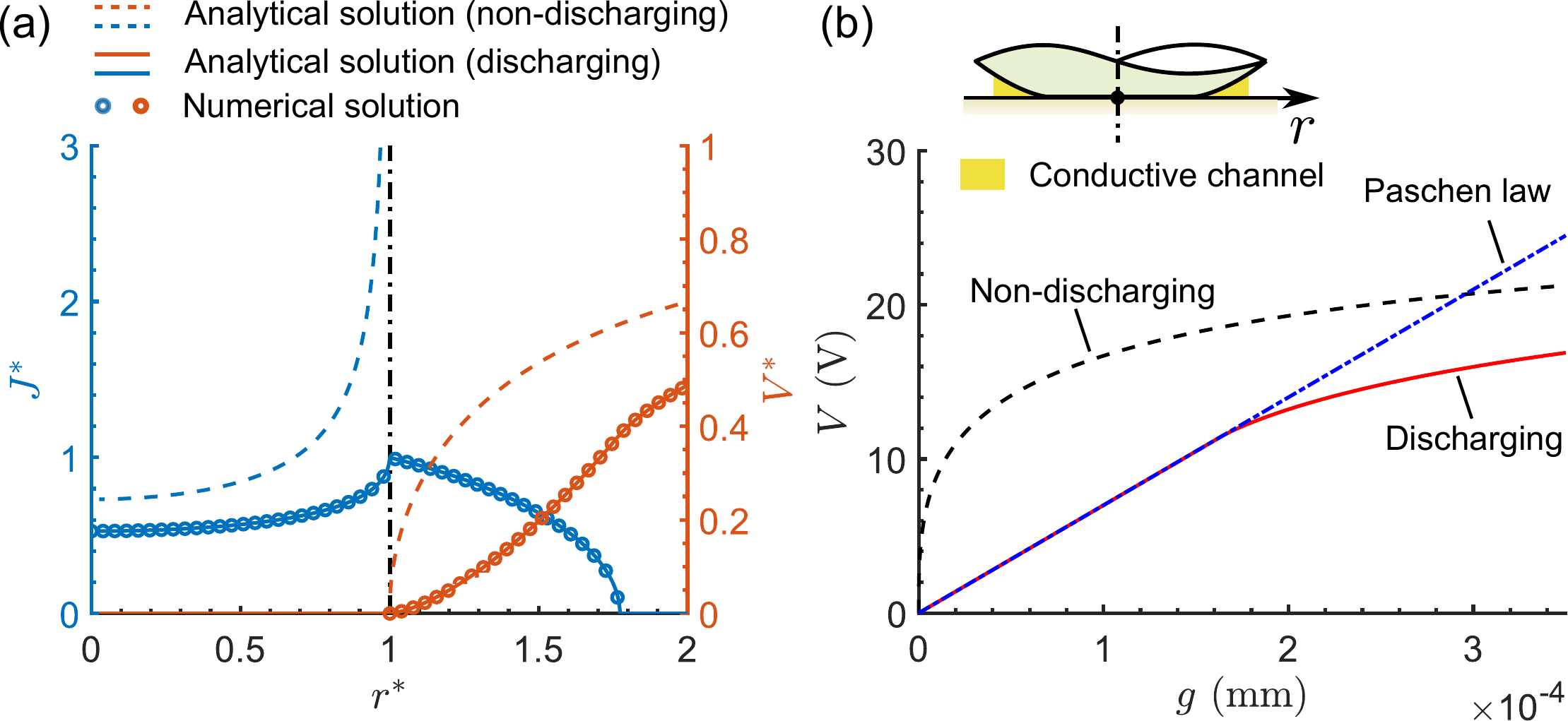}
  \caption{(a) Plots of $J^*(r^*)$ and $V^*(r^*)$ in closed forms (with and without discharging) and solved by the discharging numerical model at the contact phase with $\delta^* = 0.4667$ and $\Delta V = 30$ V. The dash-dotted line indicates the periphery of the contact area; (b) Analytical solutions of $V(g)$ with and without discharging and the modified Paschen law.}\label{fig:Fig_R1}
\end{figure}

Consider a typical electrical contact between an elastic, parabolic, copper electrode of radius $R = 0.5$ mm and a flat copper electrode. The resistivity of copper is $1.68 \times 10^{-5}$ $\Omega \cdot$m. Fig. \ref{fig:Fig_R1}(a) shows that the analytical discharging solutions at the contact phase (solid lines), $J^*(r^*)$ and $V^*(r^*)$, are nearly identical to the corresponding numerical solutions (hollow circles) with a visually indistinguishable difference. Conventional electrical contact theory predicts an inverse bell-shaped current density distribution (dashed line), where $J^*$ is fairly constant at the contact center and grows rapidly toward the contact periphery (dash-dotted line) until it becomes singular at $r^* = 1$ (see Fig. \ref{fig:Fig_R1}(a)). Since the discharge is more likely to occur adjacent to the contact periphery with a vanishing gap, the gas medium over an annulus region immediately outside the contact area becomes conductive and allows the current to flow through. The corresponding $J^*(r^*)$ distribution shown in Fig. \ref{fig:Fig_R1}(a) implies that the singular current density is canceled by discharging. Physically speaking, the conductive path induced by discharging relieves the ``traffic jam" of the current flow at the contact edge. The cancellation of the singular current density is similar to that in fracture mechanics and contact mechanics, where stress singularities at the tip of the Griffith crack \cite{griffith1921vi} and the edge of the adhesive contact \cite{johnson1971surface} are relieved, respectively, by the introduction of the yield zone \cite{dugdale1960yielding} and the cohesive zone \cite{maugis1992adhesion}. As the current concentration is resolved, $J^*(r^*)$ inside the contact area is lower than that of the non-discharging solution. As $r^* \to 1^-$, $J^*(r^*)$ grows more rapidly toward a finite peak and is followed by a monotonic drop to zero at $r^* = c/a$. The non-discharging $V^*$ monotonically increases from zero at $r^* = 1$ with a decreasing rate as $r^* (> 1)$ increases (see dashed line in Fig. \ref{fig:Fig_R1}(a)). The non-discharging $V^*$ eventually converges to unity at the distant radial location, which is not illustrated in Fig. \ref{fig:Fig_R1}(a). The discharging $V^*(r^*)$ behaves similarly to the non-discharging solution.  {It is relatively lower than the non-discharging potential with a finite slope at $r^* = 1$. The intersection between the dashed line and the modified Paschen law shown in Fig. \ref{fig:Fig_R1}(b) indicates that the potential drop associated with the small gap when a discharge is not considered is larger than the breakdown voltage governed by the modified Paschen law. This is a strong signal that discharge inevitably occurs at small gap ranges, where the $V(g)$ relation considering discharge follows almost exactly the modified Paschen law within the discharge zone. When the gap is larger than a threshold value, discharge is extinguished, and $V(r) < V_{\text{b}}(g(r))$.

\begin{figure}[h!]
  \centering
  \includegraphics[width=16cm]{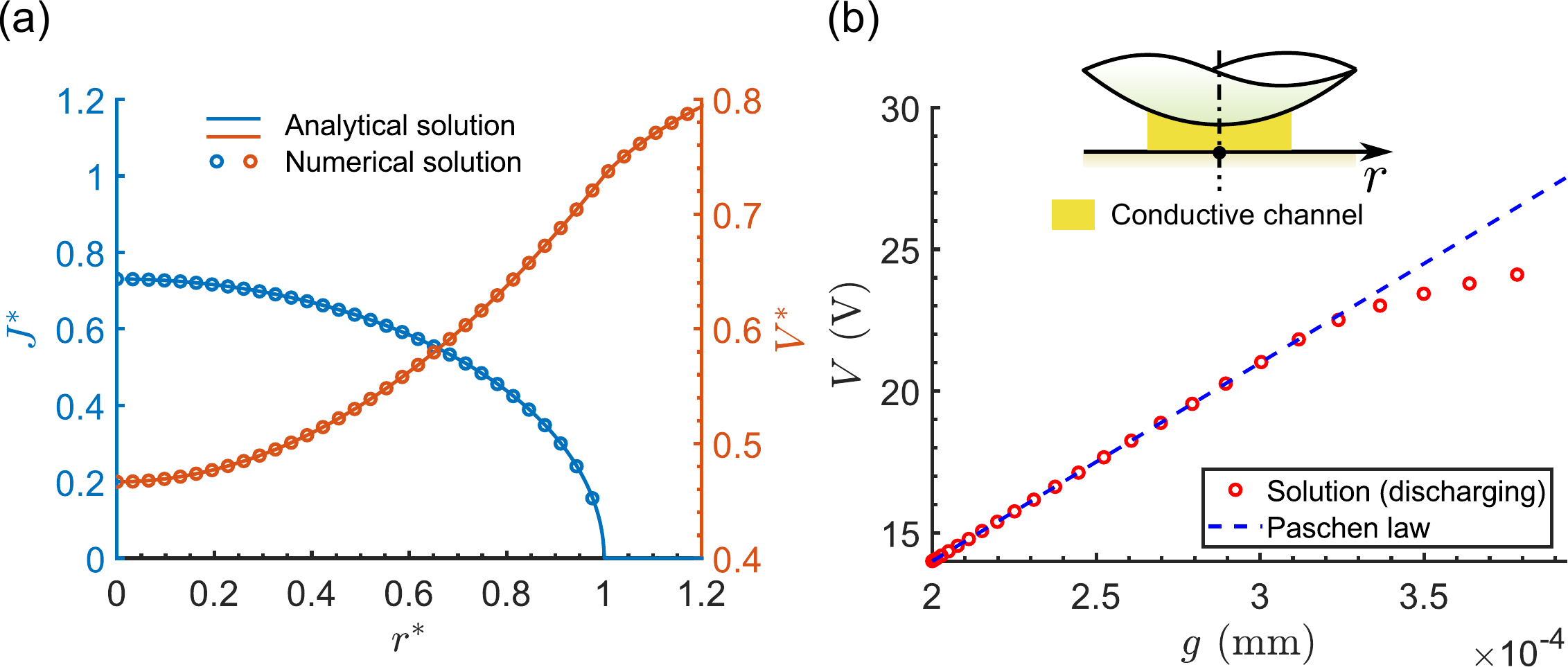}
  \caption{(a) Plots of the analytical and numerical solutions of $J^*(r^*)$ and $V^*(r^*)$at separation phase with $\delta^* = -0.4667$ and $\Delta V = 30$ V; (b) Numerical solution of the $V(g)$ relation and the modified Paschen law.  }\label{fig:Fig_R2}
\end{figure}

The validity of the analytical discharging solutions at the separation phase is confirmed by the numerical model in Fig. \ref{fig:Fig_R2}(a). The current density distribution $J^*(r^*)$ has a global maximum $r^* = 0$ and monotonically drops to zero as $r^* \to 1^-$. The corresponding $V^*(r^*)$ has a local minimum at $r^* = 0$, followed by a parabolically increasing trend within the discharge zone. The only difference between the $V(g)$ relations in the contact and separation phases is that the latter starts with a finite interfacial gap (Fig. \ref{fig:Fig_R2}(b)). Comparing the current density distribution in Fig. \ref{fig:Fig_R1}(a) and \ref{fig:Fig_R2}(a), we can expect that, as indentation transits from positive (contact phase) to negative (separation phase), the central convex current distribution gradually shrinks till vanishing. The outer concave current distribution eventually dominates the whole interface and evolves self-similarly as $-\delta$ increases.

\begin{figure}[h!]
  \centering
  \includegraphics[width=16cm]{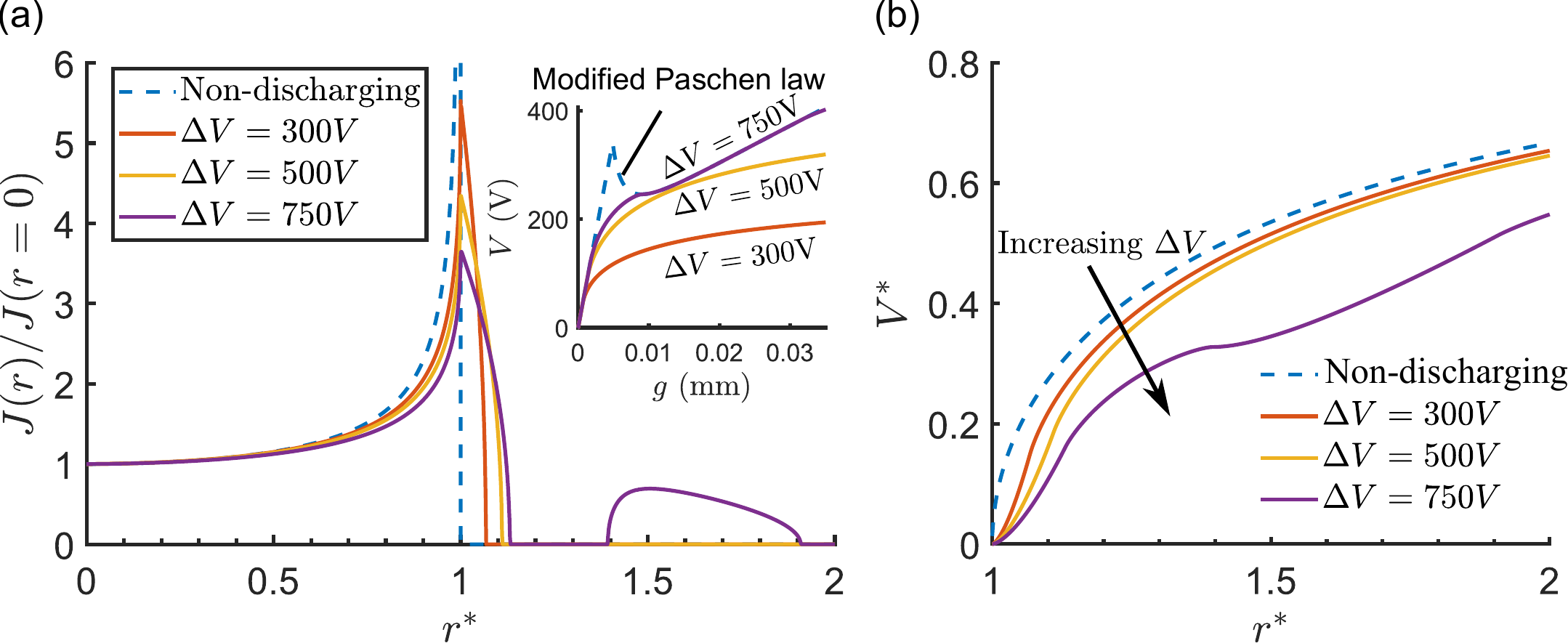}
  \caption{Plots of the numerical discharging solutions of (a) $J(r)/J(r = 0)$ and (b) $V^*(r^*)$ at contact phase with various values of $\Delta V$ and a fixed indentation $\delta = 0.0304$ mm.}\label{fig:Fig_R3}
\end{figure}

Fig. \ref{fig:Fig_R3} shows how $\Delta V$ influences $J^*(r^*)$ and $V^*(r^*)$. As $\Delta V$ increases, $J^*$ at the contact center is barely influenced by $\Delta V$, while $J^*$ greatly drops with $\Delta V$ at the contact edge. This implies that the peak-to-center ratio of the current density drops as $\Delta V$ increases. The discharge zone at the non-contact area ($r^* > 1$) gradually grows with $\Delta V$. As $\Delta V > 500$ V, the $V(g)$ relation interacts with the modified Paschen law within two distinct gap ranges (see the inset of Fig. \ref{fig:Fig_R3}(a)), and the discharge zone is split into two annulus regions ($J^*(r^*)$ with $750$ V). As $\Delta V$ further increases, two annulus regions are expected to coalesce into one. This surprising dual discharge zone is related to the non-monotonic nature of the modified Paschen law and makes it nearly impossible to obtain analytical solutions at high $\Delta V$. The piece-wise Paschen law has been confirmed both experimentally and numerically using the micro-gap breakdown tests \cite{slade2002electrical} and PIC/MCC models \cite{go2014microscale}, respectively. Thus, the dual discharge region is valid under the assumption that the micro-gaps within the discharge zone can be ``artificially" closed. Since there is a lack of physical evidence to prove the existence of this second annulus zone, future studies should focus on accurately imaging the discharge zone. Similar to the evolution of $J^*(r^*)$, $V^*(r^*)$ and its initial slope at $r^* = 1$ monotonically drops with an increasing $\Delta V$ (see Fig. \ref{fig:Fig_R3}(b)). The dimensionless potential, $V^*$, non-monotonically grows with $r^*$ temporarily within the second discharge region (see the curve with $\Delta V = 750$ V in Fig. \ref{fig:Fig_R3}(b)), and it quickly gets back to the monotonic trend as $r^*$ proceeds. 

\begin{figure}[h!]
  \centering
  \includegraphics[width=8cm]{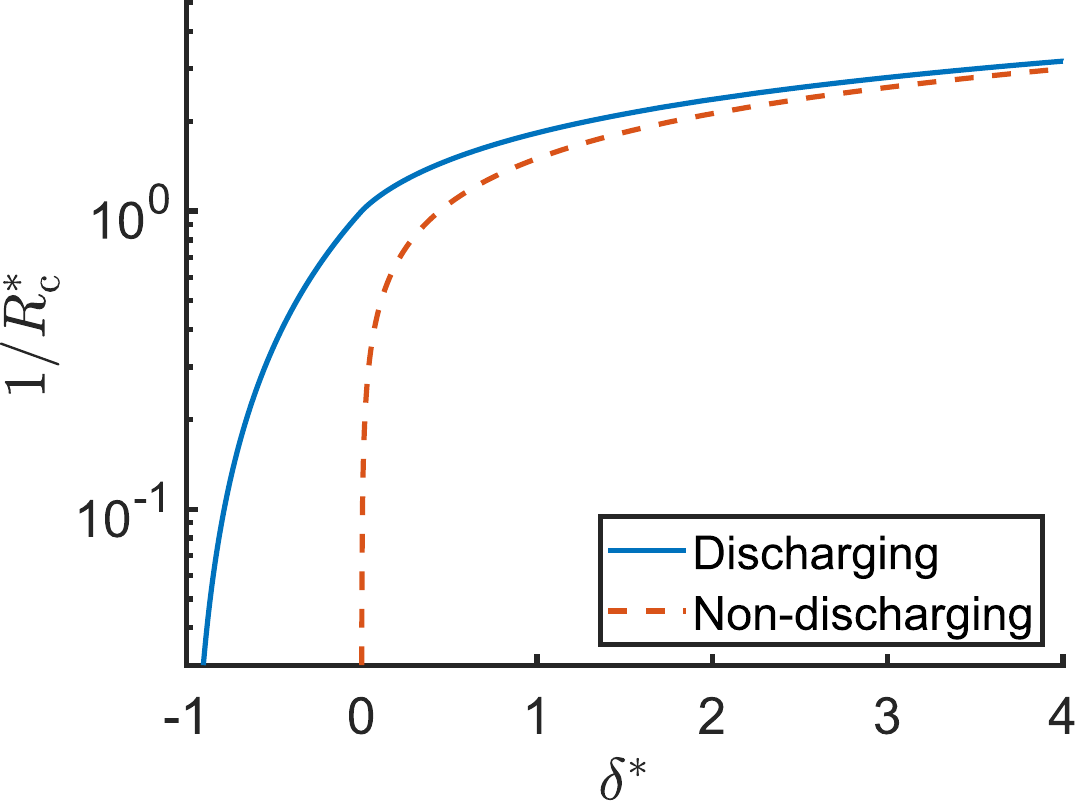}
  \caption{Variations of $1/R_{\text{c}}^*$ with $\delta^*$ in discharging and non-discharging states. }\label{fig:Fig_R4}
\end{figure}

The reciprocal of various $R_{\text{c}}^*$ (also known as the contact conductance) are shown in Fig. \ref{fig:Fig_R4}. The discharging electrical contact model provides closed-form solutions of $R_{\text{c}}^*$ at both contact and separation phases (see Eqs. \eqref{eq:Rc_dis_cont_dless} and \eqref{eq:Rc_dis_sep_dless}), and $R_{\text{c}}^*$ is continuous at $\delta^* = 0$. The discharge not only resolves the current concentration but also extends the non-singular $R_{\text{c}}^*$ to negative indentation ranges at the separation phase, and further postpones the divergence of $R_{\text{c}}^*$ from $\delta^* \to 0^+$ to $\delta^* \to -1^+$. 

Combining the single asperity electrical contact models developed in Section \ref{subsec:analalytical_solu} with various geometrical characterizations of the rough surface topography, we can include discharge in various rough surface electrical contact models \cite{greenwood1966contact,ciavarella2004electrical,kogut2004electrical,wilson2008electrical,ta2021volumetric}. Consider a simple case where an elastic, conductive, nominally flat rough surface is in electrical, purely normal contact with a conductive, rigid flat. Assuming the radii of curvature of all asperity peaks are the same, and all in-plane asperity contacts are distantly separated, constricted resistors formed by micro-contacts at the asperity level are connected in parallel. We can formulate the dimensionless electrical contact resistance, $R_{\text{c}}^*$, of the rough surface non-discharging electrical contact under the framework of the Greenwood-Williamson model \cite{greenwood1966contact}:
\begin{equation}\label{eq:ECR_nondischarge}
1/R_{\text{c}}^* = N \int_{d^*}^{\infty} \frac{3}{2} \sqrt{\delta^*} \Phi(h^*) \text{d}h^*,
\end{equation}
where $\Phi(h^*)$ is the probability density function of the dimensionless asperity peak height, $h^* = h K/\Delta V$, $d^* = d K/\Delta V$ is the dimensionless surface separation between the rigid flat and mean level of rough surface, $\delta^* = h^* - d^*$ is the dimensionless indentation, $N$ is the number of asperities over the nominal contact area. When discharge occurs at the interface, the formulation of $R_{\text{c}}^*$ is changed to
\begin{equation}\label{eq:ECR_discharge}
1/R_{\text{c}}^* = N \int_{d^*}^{\infty} \left[ (1 + \delta^*)^{3/2} - (\delta^*)^{3/2} \right] \Phi(h^*) \text{d}h^* + N \int_{d^* - 1}^{d^*} (1 + \delta^*)^{3/2} \Phi(h^*) \text{d}h^*,
\end{equation}
where the first and second integrals on the right side of Eq. \eqref{eq:ECR_discharge} are contributed by the asperities in contact and separation phases, respectively. We have shown in Section \ref{subsec:dimensionless_form} that $\frac{3}{2} \sqrt{\delta^*} < (1 + \delta^*)^{3/2} - (\delta^*)^{3/2}$, $\forall \delta^* \geq 0$. Therefore, discharging results in ECR reduction at both the asperity and rough surface levels. A similar conclusion may be deduced by using other asperity-based rough surface electrical contact models. It is also known that the ECR can be effectively estimated by the interfacial contact stiffness based on Barber's elastic-electrical analogy \cite{barber2003bounds,persson2022electric,li2024efficient}. When discharge occurs at the interface, Barber's analogy may overestimate the electrical contact resistance, and a new analogy is needed to account for discharge. 

In the numerical and analytical models (Sections \ref{subsec:numerical_model} and \ref{subsec:analalytical_solu}), the usage of the interfacial gap is conflicting: The interfacial gap is ``artificially" closed within the discharge zone so that electrons, driven by discharge events, can pass through it. However, the non-zero gap value is used to estimate the dielectric breakdown voltage based on the modified Paschen law. In practice, the discharge between two surfaces is a complex phenomenon of electron flow driven by electron impact ionization, secondary emission, field emission, etc. Particle-in-cell with the Monte Carlo collision (PIC/MCC) technique is often used in plasma dynamics to simulate the motion and interactions of the charged particles \cite{go2014microscale}. Like the other molecular/atomic dynamics models, finite computational resources restrict the application of the PIC/MCC model to a microscopic computational domain that is impractical for a macroscopic electrical contact simulation. To overcome this limitation, the PIC/MCC model may be applied locally at those discrete discharge regions to predict the outbreak of the discharge events without referring to the modified Paschen law, as well as the current density spatial distribution within the conductive channel. To further improve the efficiency of the discharging electrical contact model, future studies should focus on curve-fitting an empirical law from the numerical results of the PIC/MCC model to characterize the inter-relation between the dielectric breakdown voltage, interfacial gap, and current density within the discharge zone. Therefore, the conflicting usage of the interfacial gap is a compromising approach toward reducing the complexity of the plasma dynamics model. We can expect that, as the discharging gap is larger than the threshold value, the current density at the interface estimated by the present work gradually deviates from the numerical solutions of the PIC/MCC model. 

There is a lack of direct evidence to support the existence of single and double annulus discharge regions outside the contact area of a sphere/flat electrical contact. The EIBD in electric powertrains (e.g., rolling element bearings in electric motors, and gears), however, can serve as indirect evidence supporting the existence of the circular discharge zone when a parabolic electrode is completely separated from its mating electrode. For instance, numerous researchers \cite{janik2024exploring,bond2024influence} have experimentally observed the randomly distributed pits with various sizes and depths on the raceway of the electrified rolling element bearing. Those pits are due to the electrical corrosion induced by multiple discharge events emitted from the peaks of the asperities on the cathode, which is completely separated from the anode with a thin film lubricant. In future studies, an electrical contact test may be conducted between a cathode and a transparent anode in nitrogen gas, and discharge events may emit faint flashes of visible light through nitrogen luminescence. The shape of the discharge zone on the contact interface can be captured using a high-sensitivity electron-multiplying charge-coupled
device camera \cite{sobolev2022charge}. The image of the electrified interface shows a strong contrast between discharging and non-discharging regions, which can be efficiently segmented and binaried through image thresholding. 

The numerical model proposed in the present work can be used in conjunction with the existing boundary/mixed/elastohydrodynamic/hydrodynamic lubrication models to simulate the interfacial current density within the discharge zone, which can be further used to predict the topographical change of the lubricated surfaces caused by electrical discharge machining-like damage (e.g., frosting, pitting, fluting, etc.) \cite{he2020electrical}. In an electrified rolling element bearing, for instance, each lubricated interface can be simplified as a resistor and a capacitor in parallel. The analytical solution proposed in the present work can be used to build the electrical contact resistance and capacitance of an electrified lubricated rough interface, which serve as the foundation of the circuit model of an electrified rolling element bearing \cite{bleger2024automotive}. This work also suggests that it might be extremely difficult to completely eliminate leakage current discharge damage in rolling element bearing contacts, even if they have solid conductive contact between them. Other works have suggested that the damage only occurs when there is a non-conductive film of lubricant completely separating the surfaces.

\section{Conclusion}
In the present work, we developed a numerical model for the axisymmetric electrical contact problem where insulating air breaks down adjacent to the contact area. We derived analytical solutions for discharging electrical contacts subjected to low applied voltage in two respective scenarios: two electrodes are in contact and completely separated. We show that the singular current density at the contact periphery becomes finite once the conductive channel induced by discharging is formed adjacent to the contact area. When two electrodes are in solid-solid contact, discharge reduces the current density inside the contact area and allows electrons to flow through a much larger region. The corresponding current and ECR increases and decreases, respectively. Moreover, discharge extends the electrical contact theory to predict the current density and potential distribution between two separate electrodes. The present work theoretically proves that the ignorance of the dielectric breakdown of air can lead to singular current density at the contact periphery and ECR overestimation. It serves as a foundation for modeling discharging rough surface electrical contact and EIBD. 

\section*{Acknowledgements}
This work was supported by the National Natural Science
Foundation of China (No. 52105179), and the Fundamental Research Funds for the Central Universities (No. JZ2023HGTB0252, PA2024GDSK0044). The authors would like to thank anonymous reviewers and Dr. Haibo Zhang for their constructive comments.

\begin{spacing}{1} 
	\bibliographystyle{asmejour}
	\bibliography{ref}
\end{spacing}

\appendix
\setcounter{figure}{0}
\setcounter{equation}{0}
\renewcommand{\thefigure}{A.\arabic{figure}}
\renewcommand{\theequation}{A.\arabic{equation}}

\section*{Appendix A. Barber's elastic-electrical analogy}
Consider a general three-dimensional elastic contact problem, where two conductive half spaces are in purely normal contact. One is fixed in space, and the other is subjected to a normal load $F$. The boundaries of the two half-spaces have arbitrary topographies, $z = h_i(x, y)$, $i = 1, 2$, with negligible root mean square slopes $\sqrt{\langle |\nabla h_i|^2 \rangle} \ll 1$. The Young's modulus and Poisson's ratio are $E_i$ and $\nu_i$, $i = 1, 2$, respectively. Two constant potential boundaries, $V_1$ and $V_2 (< V_1)$, are applied respectively far from the contact. The resistivities of the two half-spaces are $\rho_1$ and $\rho_2$, respectively.

\subsection*{A.1 Elastic contact}
The interfacial gap $g(x, y)$ between two interacting surfaces is strictly zero within the contact region $\Omega_{\text{c}}$. For a non-adhesive contact, the contact pressure distribution $p(x, y)$ vanishes at the non-contact region $\Omega_{\text{nc}}$. The two interfacial properties, $p(x, y)$ and $g(x, y)$, satisfy the Karesh-Khun-Tucker (KKT) condition: 

\begin{alignat}{3}
&g(x, y) = 0, ~~~ &&p(x, y) > 0, ~~~ &&(x, y) \in \Omega_{\text{c}}, \\
&g(x, y) > 0, ~~~ &&p(x, y) = 0, ~~~ &&(x, y) \in \Omega_{\text{nc}}.
\end{alignat}
The relationship between $g(x, y)$ and $p(x, y)$ can be written as
\begin{equation}
g(x, y) = -h(x, y) + \bar{u}_z(x, y) - \delta, 
\end{equation}
where the composite rough surface $h(x, y) = \sum \limits_{i=1,2} h_i(x, y)$. The summation of the surface normal displacement of both surfaces is
\begin{equation}\label{eq:uz_bar}
\bar{u}_z(x, y) = \frac{1}{\pi E^*} \int_{-\infty}^{\infty} \int_{-\infty}^{\infty} \left[ (x - \xi)^2 + (y - \zeta)^2 \right]^{-1/2} p(\xi, \zeta) \text{d}\xi \text{d}\zeta,
\end{equation}
where the plane strain modulus $E^* = 1/[(1 - \nu_1^2)/E_1 + (1 - \nu_2^2)/E_2]$. The relative rigid body approach is $\delta$ (also known as the indentation). The two-body normal contact problem described by Eqs. (A.1--A.5) is equivalent to an indentation problem where a rigid indenter of shape $h(x, y)$ penetrating into an elastic half-space with the Young's modulus, $E$, and Poisson's ratio, $\nu$, satisfying $E^* = E/(1 - \nu^2)$. By applying an infinitesimally small indentation depth, $\Delta \delta$, the contact area is unchanged. A mixed boundary value problem consisting of the incremental change of surface normal displacement, $\Delta \bar{u}_z(x, y)$, and the corresponding contact pressure response, $\Delta p(x ,y)$, can be formulated as follows \cite{barber2003bounds}:

\begin{alignat}{3}
&\Delta \bar{u}_z(x, y) = \Delta \delta, ~~~ &&\Delta p(x, y) > 0, ~~~ &&(x, y) \in \Omega_{\text{c}}, \label{eq:elastic_cont} \\
&\Delta \bar{u}_z(x, y) < \Delta \delta, ~~~ &&\Delta p(x, y) = 0, ~~~ &&(x, y) \in \Omega_{\text{nc}}. \label{eq:elastic_ncont} 
\end{alignat}
The relation between $\Delta \bar{u}_z$ and $\Delta p$ satisfies the same convolution law given in Eq. \eqref{eq:uz_bar}.

\subsection*{A.2 Electrical contact}
Let $V^+(x, y, z \geq 0)$ and $V^-(x, y, z \leq 0)$ be the potential distribution of the two contacting half spaces. For an arbitrary interfacial current density, $J(x, y)$, flowing through the interface, the potential distributions of the two half spaces are \cite{mendez2018scaling}
\begin{equation}
V^+(x, y, z) = -\frac{\rho_1}{2 \pi} \int_{-\infty}^{\infty} \int_{-\infty}^{\infty} \left[ (x - \xi)^2 + (y - \zeta)^2 + z^2 \right]^{-1/2} J(\xi, \zeta) \text{d}\xi \text{d}\zeta + V_1,
\end{equation}
\begin{equation}
V^-(x, y, z) = \frac{\rho_2}{2 \pi} \int_{-\infty}^{\infty} \int_{-\infty}^{\infty} \left[ (x - \xi)^2 + (y - \zeta)^2 + z^2 \right]^{-1/2} J(\xi, \zeta) \text{d}\xi \text{d}\zeta + V_2,
\end{equation}
where $V_1$ and $V_2$ in the above two equations guarantee the convergence of $V^+(x, y, z)$ and $V^-(x, y, z)$ to $V_1$ and $V_2$, respectively, far from the interface. Let us define the potential drop distribution at the interface as 
\[
V'(x, y) = -\left[V^{+}(x, y, z = 0) - V^{-}(x, y, z = 0) \right] + \Delta V,
\] 
where $\Delta V = V_1 - V_2$ so that
\begin{equation}\label{eq:Vp}
V'(x, y) = \frac{\rho}{2 \pi} \int_{-\infty}^{\infty} \int_{-\infty}^{\infty} \left[ (x - \xi)^2 + (y - \zeta)^2 \right]^{-1/2} J(\xi, \zeta) \text{d}\xi \text{d}\zeta,
\end{equation}
where $\rho = \rho_1 + \rho_2$. Without the dielectric breakdown of air at the interfacial gap, the current lines can only pass through the contact region ($\Omega_{\text{c}}$), and the current density at the non-contact region ($\Omega_{\text{nc}}$) is strictly zero. The potential is continuous across the contact area and is discontinuous with a certain jump at the non-contact area. Therefore, a similar mixed boundary value problem composed of $V'(x, y)$ and $J(x, y)$ can be formulated as

\begin{alignat}{3}
&V'(x, y) = \Delta V, ~~~ &&J(x, y) > 0, ~~~ &&(x, y) \in \Omega_{\text{c}}, \label{eq:electric_cont} \\
&V'(x, y) < \Delta V, ~~~ &&J(x, y) = 0, ~~~ &&(x, y) \in \Omega_{\text{nc}}. \label{eq:electric_ncont} 
\end{alignat}

\subsection*{A.3 Elastic-electrical analogy}
The similarity between Eqs. (\ref{eq:uz_bar}--\ref{eq:elastic_ncont}) and Eqs. (\ref{eq:Vp}--\ref{eq:electric_ncont}) implies that 
\begin{equation}\label{eq:identity}
V'(x, y) = \frac{\Delta V}{\Delta \delta} \Delta \bar{u}_z(x, y), ~~~ J(x, y) = \frac{2 \Delta V}{\rho E^* \Delta \delta} \Delta p(x, y).
\end{equation}
These identities can directly deduce an important conclusion that the electrical contact resistance, $R_{\text{c}}$, is linearly dependent on the reciprocal of the contact stiffness, $K_{\perp}$, 
\begin{equation}\label{eq:Barber_identity}
R_{\text{c}} = \frac{1}{2} \rho E^* K_{\perp}^{-1},
\end{equation}
where $R_{\text{c}}$ and $K_{\perp}$ are

\begin{align}
K_{\perp} &= \iint \limits_{\Omega_{\text{c}}} \Delta p(x, y) \text{d}x \text{d}y/ \Delta \delta, \\
R_{\text{c}} &= \Delta V/\iint \limits_{\Omega_{\text{c}}} J(x, y) \text{d}x \text{d}y. 
\end{align}
Eq. \eqref{eq:Barber_identity} was originally found by Barber \cite{barber2003bounds} using harmonic potential functions. Here, we reach the same conclusion by following a slightly different approach. Barber's elastic-electrical analogy implies that solving interfacial electrical properties is equivalent to solving the incremental problem of the elastic contact (Eqs. \eqref{eq:elastic_cont} and \eqref{eq:elastic_ncont}). With some known incremental solutions of the elastic contact problem, we can directly deduce the corresponding electrical contact solutions. As an example, we will revisit electrical Hertzian contact using Barber's elastic-electrical analogy in the next section.  

\subsection*{A.4 Hertzian electrical contact}
Consider an elastic, parabolic, conductive surface in purely normal contact with an elastic, conductive, flat surface (see Fig. \ref{fig:Fig_1}(a)). The substrates beneath the two surfaces are approximately half-spaces. The solutions for the contact interface are given by Hertzian circular contact theory. The indentation, $\delta =a^2/R$, and non-zero contact pressure distribution, $p(r)$, is
\begin{equation}\label{eq:Hertzian_pressure}
p(r \leq a) = \frac{2 E^* a}{\pi R} \sqrt{1 - r^2/a^2}.
\end{equation} 
The Hertzian contact pressure distribution and indentation can be represented alternatively as \cite{hill1990concise, greenwood2010contact,liang2024general}

\begin{align}
p(r \leq a) &= \int_0^a \Delta p(r) = \frac{2 E^*}{\pi R} \int_r^a  \frac{a' \text{d}a'}{(a'^2 - r^2)^{1/2}}, \label{eq:Hertz_pressure_incremental}\\
\delta &= \int_0^a \Delta \delta  = \int_0^a \frac{2 a'}{R} \text{d}a'.
\end{align}
Therefore, Hertzian contact is equivalent to a series of circular flat-end incremental punches sequentially indenting the half-space. For an incremental punch with radius $a' \in [0, a]$ and indentation depth $\Delta \delta = 2a' \text{d}a'/R$, the incremental contact pressure distribution inside the contact area is deduced from Eq. \eqref{eq:Hertz_pressure_incremental}, i.e.,
\begin{equation}\label{eq:delta_p}
\Delta p(r \leq a') = \frac{\Delta \delta E^*}{\pi} (a'^2 - r^2)^{-1/2}, \\
\end{equation}
and the corresponding surface displacement outside the contact area is \cite{johnson1987contact}
\begin{equation}\label{eq:delta_uz_bar}
\Delta \bar{u}_z(r > a') = \frac{2 \Delta \delta}{\pi} \arcsin\left(\frac{a'}{r}\right). 
\end{equation}
Using identities in Eq. \eqref{eq:identity} and dropping the prime notation on $a$, we can obtain the closed forms of $J(r \leq a)$ and $V'(r > a)$ directly (see Eqs. \eqref{eq:J_closeform} and \eqref{eq:V_closeform}). 

\end{document}